\documentclass[review]{elsarticle}

\usepackage{lineno,hyperref}
\modulolinenumbers[5]
\usepackage[T1]{fontenc}
\usepackage{CJK}
\usepackage{xcolor}
\usepackage{amsfonts}
\usepackage{epstopdf}
\usepackage{wrapfig} 
\usepackage{subfigure}
\usepackage{graphicx}  
\usepackage{dcolumn}   
\usepackage{bm}
\usepackage{float}
\journal{Journal of \LaTeX\ Templates}
\newcommand{\be}{\begin{equation}}
\newcommand{\ee}{\end{equation}}
 \newcommand{\bea}{\begin{eqnarray}}
 \newcommand{\ena}{\end{eqnarray}}

\begin{document}

\begin{frontmatter}
\title{Thermodynamics and weak cosmic censorship conjecture of the torus-like black hole}


\author[mymainaddress]{Yi-Wen Han}

\author[mymainaddress]{Xiao-Xiong Zeng}
\cortext[mycorrespondingauthor]{Corresponding author}
\ead{xxzengphysics@163.com}
\author[mymainaddress]{Yun Hong}

\address[mymainaddress]{Department of physics, college of computer science and electronic information engineering, Chongqing Technology and Business University, Chongqing 400067, China }

\begin{abstract}

After studying the energy-momentum relation of  charged particles' Hamilton-Jacobi equations, we discuss the laws of thermodynamics and the weak cosmic censorship conjecture in torus-like black holes. We find that both the first law of thermodynamic as well as the weak cosmic censorship conjecture are valid in both the normal phase space and extended phase space. However, the second law of thermodynamics is only valid in the normal phase space. Our results show that the first law and weak cosmic censorship conjecture do not depend on the phase spaces while the second law depends. What's more, we   find that the shift of the metric function that determines the event horizon take the same form in different phase spaces, indicating that the weak cosmic censorship conjecture is independent of the phase space.
\end{abstract}

\begin{keyword}
\texttt{Black hole; Thermodynamics; Extended phase space; Weak cosmic censorship conjecture}
\MSC[2010] 00-01\sep  99-00
\end{keyword}
\end{frontmatter}

\section{Introduction}
Over the past few decades,  great progress have been made in studying  the thermodynamics of black holes. Theoretical physicists have studied the thermal and non-thermal radiation \cite{W.Hawking, Zhang, Y. Zhang, Zeng:2008zze, Zeng:2008zzc}, entropy \cite{t Hooft,Cardy}, thermodynamic phase transition \cite{Setare,Cavaglia,Chen}, and  thermodynamic properties of black holes in the frameworks of holography and geometry \cite{Zeng,Zeng.H, X. Zeng, Cai:2017ihd, W. Han, Han,Ruppeiner,G. Ruppeiner}. These results show that black hole thermodynamics is an interesting and important topic in the theory of gravity. And we know that black holes can be regarded  as thermodynamic systems with  Hawking temperature \cite{Hawking}. For the black hole thermodynamic systems, there is an irreducible mass distributed on the event horizon, which is similar to the thermodynamic entropy \cite{Christodoulou, Bardeen, Gwak:2016gwj, Gwak:2018akg, Gwak:2018tmy}. This similarity leads to the definition of  the Bakenstein-Hawking entropy, which is found to be proportional to the area of the event horizon \cite{Bekenstein}. In addition, other conserved quantities such as the thermodynamic potential of a black hole is defined at its event horizon. In other words, the event horizon is essential for  the thermodynamic system of the black hole.

In normal phase space of the thermodynamic systems, the cosmological constant that determines the spacetime asymptotic topology of the black hole is treated as a fixed parameter. However, according to the idea of Henneaux et al.\cite{Henneaux},  the cosmological constant can be treated as a dynamic variable in the thermodynamic system so that one can construct an extended phase space \cite{Caldarelli, H. Hendi, Sekiwa,Banerjee,S. Wang}. In the extended phase space, the cosmological constant $\Lambda$ was regarded as pressure $P$ \cite{Dastor} and its conjugate was regarded as volume  with the definition $V=(\partial M/\partial P)|_{S,J,Q}$.  In this case, the first law of thermodynamics for black holes was  revised to $dM=TdS+\Omega dJ+\Phi dQ+VdP$. Here, $M$ is interpreted as enthalpy \cite{Dolan,P. Dolan,David}. However, the validity of the first law of thermodynamics  in the extended phase space does not mean the second law and the  weak cosmic censorship conjecture   are valid. Thus it is interesting and important to explore the validity of the   second law and the  weak cosmic censorship conjecture   in the extended phase space of the black hole thermodynamic systems.

Recently, in the framework of extended phase space,  \cite{Gwak} investigated laws of thermodynamics and weak cosmic censorship conjecture of a charged black hole.  He found that the first law of thermodynamics was valid, and the second law was violated for extreme and near-extreme black holes. In addition, the weak cosmic censorship conjecture was found to be valid for the extremal black holes and the near-extremal black hole for they are stable. This work was then extended to the Born-Infeld gravity \cite{Zeng:2019jta} and BTZ black holes \cite{Zeng:2019jrh}, in which the contributions of other extensive quantities are considered besides the contributions  of pressure and volume.
 They also claimed that they did not employ any approximations to deal with the weak cosmic censorship conjecture.

  In this paper, we intend to investigate  laws of thermodynamics and the  weak cosmic censorship conjecture of a torus-like black hole   in the extended phase space. Different from the previous investigations, the topology of this spacetime  is $S \times S \times M^2$. We intend to explore whether the topology will affect the  weak cosmic censorship conjecture and the second law  for they have not been touched until now. We will employ the idea in
  \cite{Gwak}. As a charged particle drops into the torus-like black hole, we first obtain the energy-momentum relation near the event horizon  by solving the Hamilton-Jacobi equation. And based on this relation, laws of thermodynamics  and  weak cosmic censorship conjecture are investigated in both the normal phase space and extended phase space. We find both the first law and weak cosmic censorship conjecture are valid in the normal and extended phase space. However, the second law is violated in the extended phase space though it is valid in the normal phase space. Our result is consistent with that in  \cite{Gwak}, implying that the topology of the spacetime does not affect the thermodynamics and weak cosmic censorship conjecture.

The present work is organized as follows. In Section 2, the energy-momentum relation of the scalar particle is obtained by solving the Hamilton-Jacobi equation. In Section 3, thermodynamics and weak cosmic censorship conjecture are studied in the normal phase space. In section 4, we investigated thermodynamics and weak cosmic censorship conjecture in the extended phase space. The conclusion is presented in section 5.

\section{The dynamic behavior of charged particles in the thermodynamic system of the torus-like black hole}

The torus-like black hole solution is characterized by the line element \cite{Huang}

\begin{equation}\label{eh001}
 ds^{2}=-G(r)dt^{2}+G^{-1}(r)dr^{2}+r^{2}(d\theta^2+ d\psi^2),
\end{equation}
 where%
\begin{equation}\label{eh002}
 G(r)=-\frac{\Lambda r^2}{3 }-\frac{ 2 M}{\pi r}+\frac{4 Q^2}{\pi r^2}.
\end{equation}%
For $\Lambda< 0$, the metric has coordinate singularities at the horizon radii  $r_\pm $, which satisfy \cite{M. Sharif}
\be\label{ehyw033}
-\frac{\Lambda r_\pm^2}{3 }-\frac{ 2 M}{\pi r_\pm}+\frac{4 Q^2}{\pi r_\pm^2}=0,
 \ee
where, $r_+$ and $r_-$ correspond to  outer and inner horizons of the black hole. The  nonvanishing component of electromagnetic vector potential  is
\be\label{eh003}
A_t =-4 Q/r.
 \ee

We first focus on the dynamics behavior of a charged particle  swallowed by the black hole and its energy-momentum relationship  near the event horizon. In the electric  field, the motion of scattered particles  satisfy the Hamilton-Jacobi equation of curved space-time, which is
\begin{equation}\label{eh004}
g^{\mu \nu }\left(p_{\mu }-{eA}_{\mu }\right)\left(p_{\nu }-{eA}_{\nu }\right)+\mu_0 ^2=0,
\end{equation}
where $\mu_0$ is rest  mass of a scalar particle, $p_\mu$  the momentum of the particle, which is defined by
\begin{equation}\label{eh005}
 p_\mu=\partial_\mu \emph{S},
\end{equation}%
here, $\emph{S}$  is the Hamiltonian action. Considering the symmetry of the gravitational system, the action should be written as

\begin{equation}\label{eh006}
\emph{S} = -{Et}+ {R(r)}+{\Theta(\theta)}+ {L \psi},
\end{equation}
where, $E$  and  $L$ are the energy and momentum of the particle, and they are conserved quantities of timespace in the gravitational system. From Eq.(\ref{eh001}), the contravariant metric of the black hole  can be written as
\begin{equation}\label{eh007}
g^{\mu \nu } \partial _{\mu }\partial _{\nu }=-G^{-1}(r)\left(\partial _t\right){}^2+G(r)(\partial r)^2+ r^{-2}(\partial _\theta )^2+(r\sin\theta )^{-2}(\partial \psi )^2.
\end{equation}
According to Eq.(\ref{eh006}), the Hamilton-Jacobi equation(\ref{eh007}) can be reduced to
\begin{equation}\label{eh008}
-G^{-1}(r)\left(E +{eA}_t\right){}^2+G(r)(\partial_r R(r))^2+{r^{-2}}(\partial_\theta\Theta(\theta))^2+(r\sin\theta )^{-2}L^2+\mu_0 ^2=0.
\end{equation}
Introducing a variable $\lambda$, we are able to make variable separation of the equation (\ref{eh008}), the radial and angular components  are
\begin{equation}\label{eh009}
-r^2 G^{-1}(r) \left(-E -{eA}_t\right){}^2+r^2G (r) (\partial _r R(r))^2+r^2\mu_0 ^2=-\lambda,
\end{equation}
\begin{equation}\label{eqhy001}
(\partial_\theta \Theta(\theta))^2+\sin ^{-2}\theta L^2=\lambda.
\end{equation}
With Eqs.(\ref{eh009}) and Eq.(\ref{eqhy001}),  the radial   and the angular components of the momentum can be expressed respectively as

\begin{equation}\label{eqhy002}
p^r=G(r)\sqrt{\frac{-\mu_0  ^2r^2+\lambda}{r^2G(r)}+\frac{1}{G(r)^2}\left(-E -{eA}_t\right){}^2},
\end{equation}

\begin{equation}\label{eqqhy003}
p^{\theta }=\frac{1}{r^2}\sqrt{\lambda-\frac{1}{\sin ^2\theta }L^2}.
\end{equation}
Near the event horizon, the energy-momentum relation of charged particles is
\begin{equation}\label{eqhy004}
E=\frac{q}{r_h}e+|p^r|.
\end{equation}
The charge has nothing to do with the direction of time, it only has  relation with the interaction between the particle and the black hole. So, in order to ensure that the symbols of $E$ and $|p^r|$ are oriented towards the forward direction of time flow \cite{Christodoulou}, we will select the positive sign in front of $|p^r|$.

\section{Thermodynamics of the torus-like black hole in the normal phase space }

Now, we would like to investigate the thermodynamics of the torus-like black hole in normal  phase space.
According to the definition of surface gravity, the Hawking temperature of the black hole can be written as
\begin{equation}\label{eqhy3001}
T=\frac{-12 Q^2+3 M r_h -\pi  r_h^4 \Lambda }{6 \pi ^2 r_h^3}.
 \end{equation}
Using the relation between the Beckenstein-Hawking entropy and area, the entropy of the black hole can be expressed as
\begin{equation}\label{eqhy3002}
S={\pi}^2 r_h^2.
 \end{equation}
With Eq.(\ref{eh002}),   the relation between the mass and the event horizon of the torus-like black hole can be obtained as follows
\begin{equation}\label{eqhy3003}
M=\frac{12 Q^2-\pi  r_h^4 \Lambda }{6 r_h}.
 \end{equation}
When charged particles are scattered by the black hole, the energy and charge of the system should follow conservation laws. So, the internal energy and charge change of the black hole system should satisfy
\begin{equation}\label{eqhy333}
E =dM,\quad e=dQ,
\end{equation}
Substituting Eq.(\ref{eh003}) and Eq.(\ref{eqhy333}) into Eq.(\ref{eqhy004}), we get
\begin{equation}\label{eqyw002}
dM=\Phi dQ+p^r.
\end{equation}
Similarly, the scattered particles will affect the event horizon of the black hole, leading to the change of the metric function  $G(r)$. With the change of function $G(r)$,  $dM$ and $dQ$ can be deleted simultaneously. So, We can get  $dr_{h}$ directly, and from Eq.(\ref{eqhy3002}), the variations of the  black hole entropy can be expressed as
\begin{equation}\label{eqhy3004}
dS_h =\frac{6 \pi ^2 {p^r} r_h^3}{-12 Q^2+3 M r_h-\pi  \Lambda  r_h^4}.
\end{equation}
By using Eqs.(\ref{eqhy3001}) and (\ref{eqhy3004}), we get $T dS_h =p^r$. Therefore the system internal energy in Eq.(\ref{eqyw002}) can be rewritten as
\begin{equation}\label{eqhy3005}
dM=TdS_h+\Phi dQ,
\end{equation}
which is  the first law of the torus-like black hole thermodynamics in the normal phase space.

By using Eq.(\ref{eqhy3004}), we also can discuss the second law of thermodynamics. In normal phase space, from the Eq.(\ref{eqhy3001}), we  find $-12Q^2+3Mr_h-\pi \Lambda r_h^4 > 0 $, and put it into Eq.(\ref{eqhy3004}) we know $dS_h > 0 $.  In the case, the change of the  entropy of the black hole increases. In other words, the second law of thermodynamics for the torus-like black hole is valid in the normal phase space.

Of course, we can also test the weak cosmic censorship conjecture in normal phase space, which states that there should be a horizon to avoid the singularity to be naked. Therefore, an event horizon is needed to guarantee the validity of the weak cosmic conjecture. To do this, we want to test whether an event horizon exists when a charged particle is absorbed by the black hole. In other words, whether there is a solution for function $G(r)$.

 For the torus-like black hole, there is a minimum value  for the function $G(r)$ with the radial coordinate $r_{min}$. For the case $G(r_{min})>0$, there is not a horizon while for the case $G(r_{min})\leq0$, there are horizons always.   At $r_{min}$, the following relations should be satisfied
\bea \label{ehyw0020}
&G|_{r=r_{min}}\equiv G_{min}=\beta\leq 0,\nonumber\\
&\partial_{r}G|_{r=r_{min}}\equiv G'_{min}=0,\nonumber \\
&(\partial_{r})^2 G|_{r=r_{min}}\equiv G^{\prime\prime}_{m}>0.
\ena
For the extreme black hole, $\beta=0$.  As a charged particle is sucked by the black hole, the changes in the conserved quantities such as the mass and charge of the black hole can be written as $(M+dM, Q+dQ)$.  The locations of the minimum value and the event horizon can be written as $ r_{min}+dr_{min}$, $r_{h}+dr_{h}$ respectively. Here, we have a transformation of the function $G(r)$, which is labeled as $dG_{min}$. Using the condition $G'_{min}=0$, at $r_{min}+dr_{min}$, $G(r)$ can be expressed as
\bea \label{eqhy30050}
G|_{r=r_{min}+dr_{min}}=G_{min}+dG_{min}=\beta+\left(\frac{\partial G_{min}}{\partial M}dM+\frac{\partial G_{min}}{\partial Q}dQ\right),
\ena
 For the extremal black hole, the horizon  is located at $r_{min}$,   Eq.(\ref{eqyw002}) therefore is valid at  $r_{min}$. Substituting Eq.(\ref{eqyw002})  into  Eq.(\ref{eqhy30050}), we find  that  $dQ$ is deleted at the same time. Eq. (\ref{eqhy30050}) thus can be simplified to
\be\label{eqhy66}
G_{min}+dG_{min}=-\frac{2 p^r}{\pi  r_{min}}.
\ee
Eq.(\ref{eqhy66}) shows that $G(r_{min}+dr_{min})$  is smaller than $G(r_{min})$ when a charged particle is swallowed by the black hole. This result will cause extremal black holes to be transformed into non-extremal black holes. That is, the weak cosmic censorship conjecture is valid  in the normal phase space.

\section{Thermodynamics and  weak  cosmic censorship conjecture in extended  phase space}

 Next, we turn to investigate the thermodynamics and  weak  cosmic censorship conjecture  in extended phase space. Namely the cosmological constant $\Lambda$  will be treated as the thermodynamic pressure, and its conjugate as the thermodynamic volume, which can be expressed as respectively
\be\label{eqhy4000}
P=-\Lambda /(8 {\pi} ), V=4 {\pi}^2 \left.r_h^3\right/3.
 \ee
From Eq.(\ref{eqhy3001}), Eq.(\ref{eqhy3002}), Eq.(\ref{eqhy4000}), and $\Phi = 4 Q/r_h$, we obtain the Smarr formula
\be\label{eqhy4001}
 M=2(TS-V P)+\Phi Q.
\ee
Considering that $M$ has the physical  significance of enthalpy\cite{Dastor} in extended phase space,  it has relation to the system internal energy as
\be\label{eqhy4002}
M=U + PV.
\ee
According to the laws of conversion, when a particle is swallowed  by the black hole, the change in the conserved quantity of the system should be equal to the energy and charge of the particle. Therefore, in extended phase space,  Eq.(\ref{eqhy333}) can be rewritten as
\be\label{eqhy4003}
E=dU=d(M-PV),\quad e=dQ.
\ee
 Correspondingly, the energy in Eq.(\ref{eqyw002}) changes into
\be \label{eqhy4004}
dU=\Phi dQ+|p^r|.
\ee

Similarly, the absorbed particle must also change the event horizon of the black hole due to the backreaction. But the  event horizon is always determined by function $G(r)$. In the new horizon, $r_h+dr_{h}$, there is also a relation $G(r_h+dr_{h})=G(r_h)+dG_{h}=0$, which means
\be \label{eqhy4005}
dG_{h}=\frac{\partial G_ {h}}{\partial M}dM+\frac{\partial G_ {h}}{\partial Q}dQ+\frac{\partial G_ {h}}{\partial \Lambda}d\Lambda+\frac{\partial G_ {h}}{\partial r_ {h}}dr_ {h}=0.
\ee
With Eq.(\ref{eqhy4003}), the Eq.(\ref{eqhy4004}) can be rewritten as
\be \label{eq406}
dM-d(PV)=\Phi dQ+|p^r|.
\ee
From Eq.(\ref{eq406}), we can obtain
\be \label{eqhy4006}
d\Lambda = \frac{24 Q {dQ} -6 r_h{dM} +6 {p^r} r_h -3  \pi\Lambda  r_h^3{dr_h} }{\pi  r^4}.
\ee
Substituting Eq.(\ref{eqhy4006}) into Eq.(\ref{eqhy4005}), we can delete $dQ$ and $dM$ directly. So, there is only a  relation between $|p^r|$  and  $dr_{h}$, which  yields
\be \label{eqhy4007}
dr_h=\frac{6 {p^r} r_h^2}{-24 Q^2+6 M r_h+\pi \Lambda r_h^4  }.
\ee
Based on this relation, the variations of entropy and volume of the black hole can be expressed as
\be\label{eqhy4008}
dS_h=\frac{12 {\pi}^2 {p^r} r_h^3}{-24 Q^2+6 M r_h+\pi  \Lambda  r_h^4},
\ee
\be\label{eqhy4009}
dV=\frac{24 \pi ^2 {p^r} r_h^4}{-24 Q^2+6 M r_h+\pi \Lambda r_h^4  }.
\ee
Using the above formulas, we find $T dS_h-PdV=|p^r|$.  The internal energy in Eq.(\ref{eqhy4004}) thus would change into
\be  \label{eqhy4100}
dU=\Phi dQ + T dS_h-PdV.
\ee
 With Eq.(\ref{eqhy4002}), we can obtain the relation between the enthalpy and internal energy in extended phase space, that is
\be  \label{eqhy4101}
dM=d U+P dV+VdP.
\ee
Substituting Eq.(\ref{eqhy4101}) into Eq.(\ref{eqhy4100}), we get
\be\label{eqhy4102}
dM=TdS_h+\Phi dQ+VdP,
\ee
which is the first law in extended phase space. That is, in the extended phase space, the first law of thermodynamics still holds as a  charged particle is absorbed by the black hole.

We also can discuss the second law of thermodynamics in extended phase space. We would   consider the case of extreme black holes firstly. By using the condition $T =0$, we can obtain $r_h$ from Eq. (\ref{eqhy3001}) and substitute $r_h$ into Eq.(\ref{eqhy4008}), the change of the black hole entropy is given by
\be \label{eqhy4104}
dS_{extremal}=\frac{4 \pi p^r}{\Lambda r_h}.
\ee
In   Eq.(\ref{eqhy4104}), we know  ${dS_h<0}$ for ${\Lambda < 0}$. This result shows that change of the entropy of the torus-like black hole is always decreased. That is, the second law of thermodynamics is violated for the extremal black hole.

For the non-extremal black holes, we can get $Q$ with Eq.(\ref{eh002}), that is
\be \label{zeng}
Q=\frac{\sqrt{r_h} \sqrt{6 M+\pi  r_h^3 \Lambda }}{2 \sqrt{3}}.
\ee
Substituting this equation into Eq.(\ref{eqhy4008}), we find
\be
dS_h=-\frac{12 \pi ^2  p^r  r_h^2}{6 M+\pi  r_h^3 \Lambda }.
\ee
It is obvious that $dS_h$ is negative for we know $6 M+\pi  r_h^3 \Lambda$ is positive from  Eq.(\ref{zeng}). The second law of thermodynamics is also violated for the non-extremal black hole.

Similarly, we also can examine the weak cosmic censorship conjecture in extended phase space. Considering the backreaction, and   using Eq.(\ref{ehyw0020}), the conserved quantity such as mass $M$, charge $Q$, and  cosmological constant $\Lambda$ will change into $(M+dM, Q+dQ, \Lambda+d\Lambda)$  as a charged particle  drops into the black hole.  The locations of the minimum value and the event horizon will then change into $ r_{min}+dr_{min}$, $ r_{h}+dr_{h}$. At $r_{min}+dr_{min}$, the change of function $G(r)$ should satisfy
\be \label{eeqhy5000}
G|_{r=r_{min}+dr_{min}}=(\frac{\partial G_{min}}{\partial M}dM+\frac{\partial G_{min}}{\partial Q}dQ+\frac{\partial G_{min}}{\partial \Lambda}d\Lambda).
\ee
For the extremal black hole, the event horizon  is located at $r_{min}$,  Eq.(\ref{eqhy4004}) thus is applicable at  $r_{min}$.  Substituting Eq.(\ref{eqhy4004}) into Eq.(\ref{eeqhy5000}), we get
\be \label{eeqhy5001}
G_{min}+ dG_{min}= -\frac{2 {p^r}}{\pi {r_{min}}}+\Lambda {r_{min}}{dr_{min}} .
\ee
 By using $G'_{min}=0$, we can get the mass of the black hole, and further
\be\label{zeng1}
dM= \frac{4  dQ  Q}{r_{min}}-\frac{1}{6}  d\Lambda   \pi  r_{min}^3+dr_{min} \left(-\frac{2}{3} \pi  r_{min}^2 \Lambda -\frac{12 Q^2-\pi  r_{min}^4 \Lambda }{6 r_{min}^2}\right).
\ee
In addition, according to $f(r_h=r_{min})=0$, we can obtain
\be\label{zeng2}
dr_{min}=-\frac{r_{min} \left(24 dQ Q+d\Lambda \pi  r_{min}^4\right)}{3 \left(-4 Q^2+\pi  r_{min}^4 \Lambda \right)}.
\ee
At the new lowest point, there is also a relation
\be
\partial_{r} G|_{r=r_{min}+dr_{min}}=G'_{min}+dG'_{min}=0.
\ee
which means
\be \label{eeqhy50}
dG'_{min}=\frac{\partial G'_{min}}{\partial Q}dQ+\frac{\partial G'_{min}}{\partial \Lambda}d\Lambda+\frac{\partial G'_{min}}{\partial r_{min}}dr_{min}=0.
\ee
From  Eqs.(\ref{eq406}), (\ref{zeng1}), (\ref{zeng2}), (\ref{eeqhy50}), we know that there are four equations with respect to $dM, dQ, dr_{min}, d\Lambda$. We can get $dr_{min}$ by solving them. Interestingly, we find $dr_{min}=0$. Thus  Eq.(\ref{eeqhy5001}) can be simplified as

\be \label{eeqhy50001}
G_{min}+ dG_{min}= -\frac{2 {p^r}}{\pi {r_{min}}},
\ee
 which is the same as that in Eq.(\ref{eqhy66}). That is, as the contribution of the  $PV$ term is considered, the  weak cosmic censorship conjecture is also valid. Interestingly, the shift of the metric function that determines the event horizon take the same form  in both the normal phase space and extended phase space.

\section{Conclusion}
In our work, by solving the Hamilton-Jacobi equation, we first obtained the energy-momentum relation near the horizon as a charged particle drops into the torus-like black hole.
 Then we got the first law of thermodynamics in both the normal and extended phase space. It should be noted that though some work \cite{P. Dolan,David} have studied the thermodynamics of a class of black holes in extended phase space. However, their work is premised on the existence of the first law of thermodynamics. While in this paper, we investigated the dynamics of the scalar particles under the scattering of black holes.   Our work  proves the rationality of the previous work \cite{P. Dolan,David}.

 The validity of the first law of thermodynamics  doesn't mean the second law is valid. Therefore, we   investigated the change  of the entropy of the extremal and non-extremal  black holes in the  normal and extended phase space. We found that the change  of entropy  for both the extremal and non-extremal  black holes were  positive in the normal phase space while negative in the extended phase space, which indicates that the second law is valid in the normal phase space and violated in the extended phase space. The reason causing the difference stem from the contribution of pressure and volume, though the physical mechanism is not clear.

  Finally, we investigated the weak cosmic oversight hypothesis by examining the metric function that determines the horizon of the black hole at the minimum point in both the normal and extended phase space.
  As a charged particle is swallowed, we found that the changes of the metric function were negative in both the normal and extended phase space. In this case, there are always solutions for the torus-like black hole so that the singularity of the spacetime is hidden, the  weak cosmic censorship conjecture thus is valid always.  Interestingly, we found that  the shift of the metric function  take the same form  in both the normal phase space and extended phase space, which indicates that the   weak cosmic censorship conjecture  is independent of the phase space.  The extremal black holes in both phase space thus would change into non-extremal black holes. Our result is different from that  in \cite{Gwak}, where the author found that the extremal black holes would be extremal black holes in the extended phase space.  The reason maybe be from that  we does not employ any approximation while  \cite{Gwak} employed.

\section*{Acknowledgements}{This work is supported  by the National
Natural Science Foundation of China (Grant Nos. 11875095), and Basic Research Project of Science and Technology Committee of Chongqing (Grant No. cstc2018jcyjA2480).}


\end{document}